\documentclass[aip,amsmath,amssymb]{revtex4-1}

\usepackage{graphicx}% Include figure files

\draft % marks overfull lines with a black rule on the right

\begin{document}

%\preprint{}

\title{Extreme matter compression caused by radiation cooling effect in gigabar shock wave driven by laser-accelerated fast electrons} %Title of paper



\author{Gus'kov,~S.~Yu.}
\affiliation{P.N.~Lebedev Physical Institute of Russian Academy of Sciences}
\author{Kuchugov,~P.~A.}
\email[]{Corresponding author: Kuchugov P.A., pkuchugov@gmail.com}
\affiliation{P.N.~Lebedev Physical Institute of Russian Academy of Sciences}
\affiliation{Keldysh Institute of Applied Mathematics of Russian Academy of Sciences}
\author{Vergunova,~G.~A.}
\affiliation{P.N.~Lebedev Physical Institute of Russian Academy of Sciences}


\date{\today}

\begin{abstract}
 Heating a solid with laser-accelerated fast electrons is unique way for a laboratory experiment to generate a plane powerful shock wave with a pressure of several hundred or even thousands of Mbar. Behind the front of such a powerful shock wave, dense plasma is heated to a temperature of several keV. Then, a high rate of radiation energy loss occurs even in low-$Z$ plasmas. The effect of strong compression of matter due to radiation cooling in a gigabar shock wave driven by fast electrons is found in computational and theoretical researches. It is shown that the effect of radiation cooling leads to the compression of matter in the peripheral region of shock wave to a density several times larger than the density at its front. Heating a solid by a petawatt flux of laser-accelerated fast electrons allows one to surpass the gigabar pressure level of a plane shock wave, which is the maximum level for the impact of laser-accelerated pellets. Higher pressure about 100~Gbar can be achieved under laboratory conditions only when a spherical target is imploded under the action of a terawatt laser pulse.
\end{abstract}


\maketitle %\maketitle must follow title, authors, abstract and \pacs

\section{Introduction}\label{sec:001}

Heating a substance with laser-accelerated charged particle beam is an effective way to generate a plane powerful shock wave with a pressure of several hundred or even thousands of Mbar~\cite{Guskov2012,Ribeyre2013,Guskov2014} in a laboratory experiment. This is due to the fact that the energy flux density of such beam is close to the intensity of laser pulse that produces it. At the same time, in contrast to laser radiation, charged particles transmit their energy in Coulomb collisions and are able to heat a dense substance with a density that significantly exceeds the critical plasma density. The permanent energy growth of modern laser facilities with terawatt and petawatt power allows us to consider laser-accelerated charged particle beam as an effective tool for generating the super-powerful shock waves in sufficiently large volume of matter that meet the needs of such important applications as inertial confinement fusion (ICF), study of matter equation of state (EOS) and laboratory astrophysics. The above primarily applies to laser-accelerated electron beam, since the efficiency of laser energy conversion into fast electron energy is significantly (2-3 times) larger than into fast ions.

Heating a solid by a petawatt flux of laser-accelerated fast electrons is the most effective method for generating a plane shock wave with the extreme pressure for a laboratory experiment. Its potential capabilities exceed ones of the method of impact of laser-accelerated pellets~\cite{Cauble1993,Karasik2010} which, in a modern experiment provided the generation of a plane shock wave with a record pressure of 740~Mbar~\cite{Cauble1993}. This record pressure is close to the maximum achievable one when using the impact method, since the collisional mechanism of laser radiation absorption is limited by the value of the coupling parameter $I \lambda^2 \approx 2 \cdot 10^{14}$~W~$\mu$m$^2$/cm$^2$ ($I$ and $\lambda$ are, respectively, the intensity and wavelength of laser radiation). The record pressure was determined by the limiting intensity for the third Nd-laser harmonic of $10^{15}$~W/cm$^2$. The method of direct heating of a solid by laser-accelerated fast electrons is designed to use intensities exceeding the collisional absorption limit, when a significant fraction of laser energy is transformed into the energy of fast electrons. It was shown in~\cite{Guskov2020} that the use of a laser pulse with an intensity of $I \approx 10^{19}-10^{21}$~W/cm$^2$ to heat a solid by fast electrons can provide the generation of plane shock wave with a pressure of several tens of Gbar. A higher pressure, up to several hundred Gbar in a laboratory experiment, can be achieved only in the case when a spherical target is imploded under the impact of a terawatt laser pulse. Therefore, to study, for example, the equation of state of matter, the method of heating of solid by laser-accelerated fast electrons is very promising, bearing in mind that diagnostics in a spherical experiment turns out to be a more difficult task in comparison with those traditional methods that can be applied in experiments with a plane shock wave.

This work is devoted to further study of the properties of a shock wave driven by heating a substance with laser-accelerated electron beam. In~\cite{Guskov2020}, it was shown that in this heating method, the radiation energy loss is the main factor limiting the temperature of produced plasma. In the compressed material behind the shock wave front, radiation energy loss plays an even greater role than in the heated region. The effect of extreme matter compression due to radiation cooling in a shock wave driven by laser-accelerated fast electrons is found on the basis of computational and theoretical researches. The effect of matter compression increase due to radiation cooling is well known in relation to laminar flows under $Z$-pinch~\cite{Pease1957,Vikhrev1978,Bernal2002,Craxton2015} and ICF target~\cite{Craxton2015,Golovkin2006} implosions. It should be noted that the radiation cooling effect is a central problem in astrophysics and laboratory astrophysics, in particular, in the sections of accretion physics~\cite{Blondin1989} and radiative laboratory shock physics~\cite{Laming2004}. In this paper the effect of radiation cooling behind the front of a powerful shock wave is considered. It is shown that the radiation cooling leads to compression of matter in the peripheral region of shock wave to a density several times larger than the density at its front. First it is discussed the features of the radiation cooling effect in a powerful shock wave driven by heating a dense substance with laser-accelerated electrons. Then the results of numerical calculations are presented and discussed.

\section{The features of shock wave driven by laser-accelerated fast electrons}\label{sec:002}

It is considered the generation and propagation of shock wave driven by heating the boundary region of semi-space with monoenergetic laser-accelerated fast electron flow. The dimensional parameters of the problem are the energy flux density of fast electrons $I_h$, the mass range of heating particles in heated substance $\mu_h$, which is a function of the initial fast electron energy $\varepsilon_h$ and the density of substance $\rho_0$. According to numerous experiments and theoretical models, the laser energy conversion into fast electron energy $\eta = I_h / I_L$ lies in the range 0.1-0.3. Despite this relatively low conversion efficiency, laser-accelerated electrons are the undisputed record holder for energy flux density among charged particle beams of any other laboratory origin, taking into account the laser intensity 10$^{20}$--10$^{21}$~W/cm$^2$ achieved in the modern experiment. The energy $\varepsilon_h$ increases with laser intensity $I_L$ and wavelength $\lambda$ and can reach ultrarelativistic values. The dependence $\varepsilon_h\left(I_L, \lambda\right)$ is given by well-known scalings~\cite{Beg1997,Haines2009}, which combine the data from numerous experiments and theoretical models
\begin{equation}\label{eq:001}
    \varepsilon_h[\mbox{MeV}] = \left\{
        \begin{array}{l}
            0.45 \left(I_{L(19)} \lambda_{\mu}\right)^{1/3},\; I_{L(19)} \lambda_{\mu}^2 < 0.1 \;, \\
            1.2 \left(I_{L(19)} \lambda_{\mu}\right)^{1/2}, \; I_{L(19)} \lambda_{\mu}^2 > 0.1 \;,
        \end{array}
    \right. \,
\end{equation}
where $I_{L(19)}$ and $\lambda_{\mu}$ are measured in 10$^{19}$~W/cm$^2$ and $\mu$m.

The calculated results were obtained for the case of impact of laser irradiation on aluminum, which is often used as a reference material in EOS experiments. For estimates, an approximation formulas are used for mass ranges of non-relativistic and relativistic electrons, which are calculated in accordance with the data of~\cite{Ribeyre2013,Atzeni2008,Honrubia2006} for aluminum plasma with an ion charge $Z = 11$:
\begin{equation}\label{eq:003}
    \mu[\mbox{g/cm$^2$}] \approx \left\{
        \begin{array}{l}
            0.8 \varepsilon_h^2, \; \varepsilon_h < 1~\mbox{MeV} \\
            0.9 \varepsilon_h, \; \varepsilon_h > 1~\mbox{MeV}
        \end{array}
    \right. \,
\end{equation}
where energy $\varepsilon_h$ is measured in MeV.

The scales of physical quantities of the problem are determined by the thermodynamic parameters of a region heated by fast electrons. In plane geometry, the mass of  heated layer remains constant and equal to the fast electron mass range, despite the increase in the heated layer temperature and the decrease in its density due to thermal expansion. Under these conditions, the thermodynamic state of heated layer is described by the solutions of Ref.~\cite{Guskov2012} for a period of quasi-static heating, when the motion of the substance can be ignored
\begin{equation}\label{eq:004}
    \rho = \rho_0, \; T = T_h \frac{t}{t_h}, \; P = P_h \frac{t}{t_h}, \; 0 \le t \le t_h
\end{equation}
and for the subsequent period of thermal expansion of the layer with a constant mass
\begin{equation}\label{eq:005}
    \rho = \rho_0 \left(\frac{t_h}{t}\right)^{3/2}, \; T = T_h \frac{t}{t_h}, \; P = P_h \left(\frac{t_h}{t}\right)^{1/2}, \; t \ge t_h.
\end{equation}
In these expressions $t_h$ is the duration of quasi-static heating period or the time of ablation loading (following to the notation of Ref.~\cite{Guskov2012}), during which isothermal rarefaction wave propagates from the outer surface of semi-space to the inner boundary of the heated layer:
\begin{equation}\label{eq:006}
    t_h \equiv \frac{\mu}{2 \rho_0 c_T} = \frac{1}{2} \left[\frac{9}{2 \left(\gamma - 1\right)}\right]^{1/3} \frac{\mu}{\rho_0^{2/3} I_h^{1/3}}.
\end{equation}
$c_T = \left(C_V \left(\gamma - 1\right) T_h\right)^{1/2}$ is the isothermal sound speed in the heated region, $T_h$ and $P_h$ are temperature and pressure that are reached at the end of the quasi-static heating period
\begin{equation}\label{eq:007}
    T_h = \frac{1}{C_V} \left[\frac{9}{16 \left(\gamma - 1\right)}\right]^{1/3} \left(\frac{I_h}{\rho_0}\right)^{2/3}, \; P_h = \left[\frac{3 \left(\gamma - 1\right)}{4}\right]^{-2/3} \rho_0 \left(\frac{I_h}{\rho_0}\right)^{2/3},
\end{equation}
$I_h$ is flux energy density of fast electrons, $\mu_h$ is mass range of the electron with energy $\varepsilon_h$, $C_V = (Z + 1) k_B / \left(A (\gamma - 1) m_p\right)$ is specific heat at constant volume, $k_B$ is the Boltzmann's constant, $m_p$ is the proton mass, $A$ and $Z$ are the atomic number and ion charge, $\gamma$ is specific heats ratio.

Using expressions~(\ref{eq:001}) and~(\ref{eq:003}) for energy and mass range of fast electron, it is easy to get that the characteristic time of thermodynamic state evolution  $t_h$ increases with both the laser intensity and wavelength. In the non-relativistic case $t_h$ grows as $t_h \sim I_L^{1/3} \lambda^{4/3}$ and in the relativistic case it grows as $t_h \sim I_L^{1/6} \lambda$. In aluminum plasma at $I_L = 10^{18}$~W/cm$^2$, $\lambda = 1.06$~$\mu$m (the 1$^{\mbox{st}}$ harmonic of the Nd-laser), that corresponds to $\varepsilon_h = 200$~keV, and at the conversion $\eta = 0.2$, the ablation loading time is about 130~ps.

From the point of view of achieving an extreme state of matter, the most interesting is the initial period of shock wave propagation, when a quasi-static heating occurs. Then, the characteristics of shock wave can be determined in the approximation of a uniform pressure distribution in the heated region and in the region involved into motion by the shock wave. Then, the velocity of shock wave $D_{SW}$ and the temperature $T_{SW}$ behind its front are expressed in terms of thermodynamic parameters of the heated region as
\begin{equation}\label{eq:008}
    D_{SW} \approx \left(\frac{\gamma + 1}{2}\right)^{1/2} c_T, \; T_{SW} \approx \frac{P_h}{C_V \rho_{SW}} = \left(\frac{\gamma - 1}{\gamma + 1}\right) T_h.
\end{equation}

The scales of pressure $P_h$ and temperature $T_h$ during the quasi-static heating depend only on the flux heating energy density $I_h$ and do not depend on the heating particle energy $\varepsilon_h$. This is a fundamental difference between substance heating and shock wave generation driven by charged particle beam and laser pulse. The pressure and temperature of laser-heated substance depend on the energy of a light quantum through the value of critical plasma density $\rho_{cr}$, which is scale of density in the region of absorption of radiation with the given quantum energy~\cite{Afanasiev1993,Lindl1995}: $T_L \sim \left(I_L / \rho_{cr}\right)^{2/3}$, $P_L \sim \rho_{cr}^{1/3} I_L^{2/3}$ ($\rho_{cr} \approx 1.83 \cdot 10^{-3} A / Z \lambda_{\mu}^2$~g/cm$^{3}$, $\lambda_{\mu}$ is laser radiation wavelength measured in $\mu$m).

At the same values of $I_h$  and $I_L$, the pressure of plasma heated by fast electrons is, approximately, by factor $(\rho_0 / \rho_{cr})^{1/3}$ larger and the temperature, in contrary, by factor $(\rho_0 / \rho_{cr})^{2/3}$ smaller in comparison with the case of laser heating. However, in a shock wave, the temperature ratio is reversed. Indeed, in the approximation of a uniform distribution of pressure, the ratio of temperatures in shock waves driven  by fast electron and laser radiation is $T_{SW(h)} / T_{SW(L)} = P_h/ P_L = (\rho_0 / \rho_{cr})^{1/3}$. The temperature in shock wave driven by powerful laser pulse is, as usual, several tens of eV, whereas the temperature in shock wave driven by fast electron beam can reach the value of several keV. Such a high temperature is a distinctive feature of the shock wave driven by fast electrons. Generation of shock wave with a Gbar-pressure and keV-temperature is a record opportunity for a laboratory experiment. Such a large temperature of dense plasma is the reason for the intense bremsstrahlung emission and the associated effect of increasing of matter compression in the transparent plasma of shock wave. Thermal conductivity takes place at the ablation boundary (in the region of a shock wave piston) and has almost no smoothing effect in a shock wave region. Thus, the conditions arise for radiation cooling and, as a result, for increase in the compression of matter in shock wave. Simple estimates relating to the quasi-static heating confirm this fact. The averaged mean free path of thermal radiation in the region involved into shock wave could be estimated as~\cite{Zeldovich1967}
\begin{equation}\label{eq:009}
    l_r = \frac{4 \sigma T^4}{W_r}
\end{equation}
where $\sigma = 1.03 \cdot 10^{17}$~J $\cdot$ cm$^{-2}$ $\cdot$ s$^{-1}$ $\cdot$ keV$^{-4}$ is Stefan-Boltzmann constant, $W_r$ is emissivity of plasma electrons~\cite{Zeldovich1967}
\begin{equation}\label{eq:010}
    W_r = 1.73 \cdot 10^{17} \left(\frac{Z}{A}\right)^2 Z T^{1/2} \rho^2, \; \mbox{J $\cdot$ cm$^{-3}$ $\cdot$ s$^{-1}$}
\end{equation}
temperature $T$ and density $\rho$ are measured in keV and g/cm$^3$, respectively.

Below it is considered an example that corresponds to irradiation of aluminum target with the 1$^{\mbox{st}}$ harmonic Nd-laser radiation with an intensity of $10^{18}$~W/cm$^2$ ($\varepsilon_h = 200$~keV) at the conversion $\eta = 0.2$, which corresponds to the fast electron energy flux density $2 \cdot 10^{17}$~W/cm$^2$. According to~(\ref{eq:003}), the mass range of fast electron with energy of 200~keV in an aluminum plasma with a charge $Z = 11$ is about of 0.032~g/cm$^2$. Further, according to~(\ref{eq:006})-(\ref{eq:008}) at $\gamma = 5/3$, the duration $t_h$ of quasi-static heating and the average temperature for this period behind the shock wave front are about 130~ps and 1.5~keV. Substituting $T = 1.5$~keV, $\rho \approx 4 \rho_0 = 10.8$~g/cm$^3$ and $Z = 11$ in~(\ref{eq:009}) and~(\ref{eq:010}), for the radiation mean free path we get the value $l_r \approx 0.03$~cm, which is about 10 times larger the size of shock wave region  $l_{SW} = D_{SW} t_h \approx 0.004$~cm. Then, for the considered example, the flux energy density of thermal radiation carried away, $q_r = W_r l_{SW}$, is about of $10^{17}$~W/cm$^2$, which is half of fast electron energy flux density. In the approximation of adiabatic compression of a substance in a radiation-cooled region the increase in density in this region compared to the density at the shock wave front can be estimated as
\begin{equation}\label{eq:011}
    \frac{\rho}{\rho_{SW(0)}} \approx \left(1 - \frac{q_r}{q_h}\right)^{-1/\left(\gamma - 1\right)}
\end{equation}
where $\rho_{SW(0)} = \rho_0 \left(\gamma + 1\right) / \left(\gamma - 1\right)$. This estimate for $q_r / q_h = 0.5$ gives an excess density in the radiation-cooled region compared to the density at the shock wave front of about 2.8. This means that a density of about 30~g/cm$^3$ can be achieved in an aluminum target.

Below it is discussed plasma heating by a beam of laser-accelerated fast electrons from the point of view of the features of the formation of the thermodynamic state of the resulting plasma. With a deceleration length of a sub-relativistic fast electron near 100-200~$\mu$m, the time for transferring its energy to plasma electrons is about 1~ps. This time is much shorter than the hydrodynamic time of the problem, which is about 100-200~ps; a fast electron transfers its energy to a plasma with stationary distributions of thermodynamic parameters. In turn, the energy transfer time is much longer than the plasma relaxation times -- the electron-electron and electron-ion energy relaxation times, which for a plasma density in the heating region of several g/cm$^3$ and a temperature of several keV are 0.0001~ps and 0.1~ps, respectively. The collisional ionization time is about 0.001~ps and occurs with the participation of plasma electrons with a Maxwellian spectrum. The main recombination mechanism under the conditions of the problem under consideration is triple collisions. The recombination rate in triple collisions exceeds the photorecombination rate by more than 100 times, which means the establishment of the Saha equilibrium in terms of the ionization composition. Thus, the heating of the substance by laser-accelerated fast electrons occurs while maintaining the local thermodynamic equilibrium of the plasma.

\section{Numerical results and discussion}\label{sec:003}

\begin{figure}[!ht]
    \centering
    \begin{minipage}[b]{0.45\textwidth}
        \includegraphics[width=\textwidth]{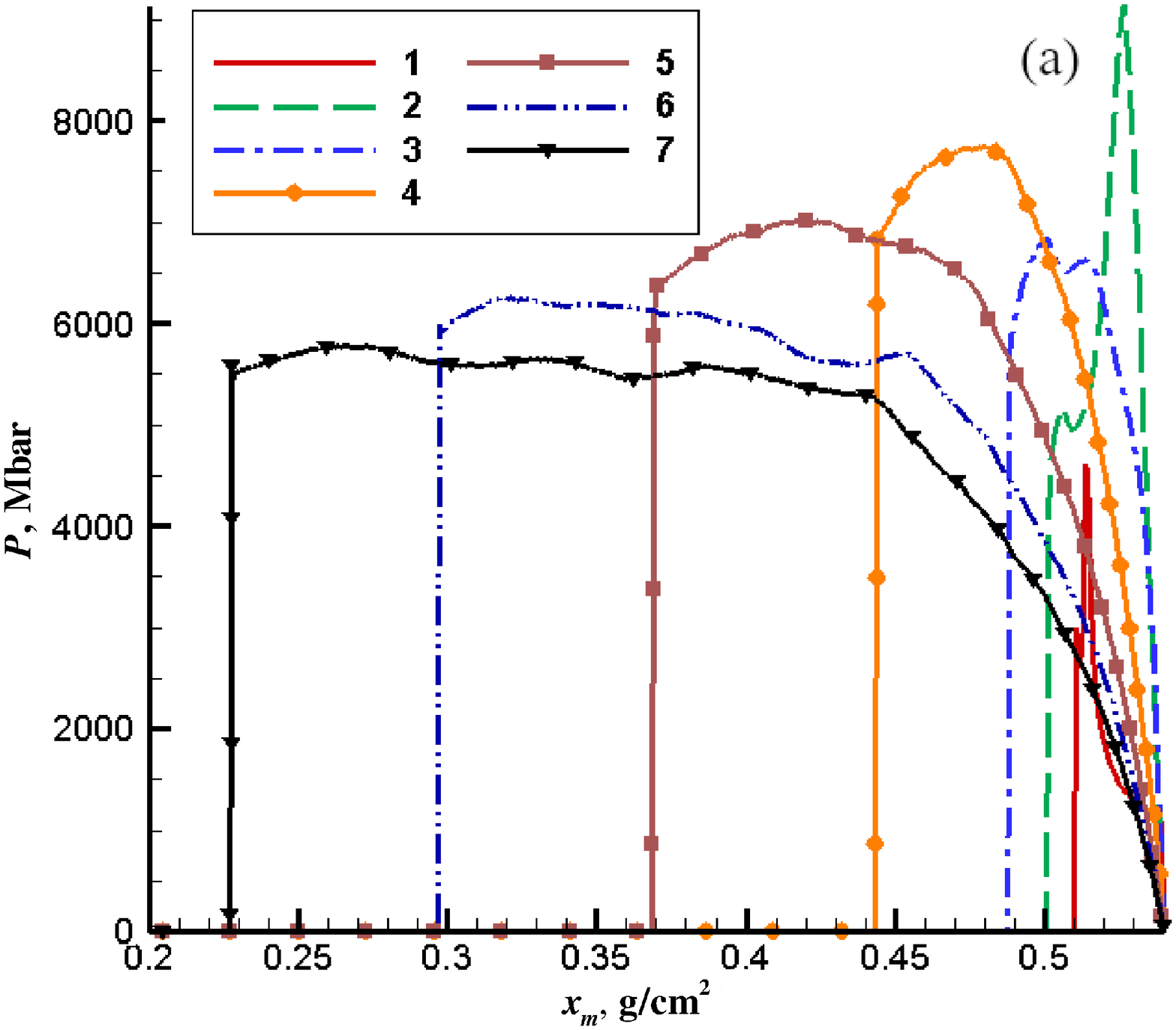}
    \end{minipage}
    \quad
    \begin{minipage}[b]{0.45\textwidth}
        \includegraphics[width=\textwidth]{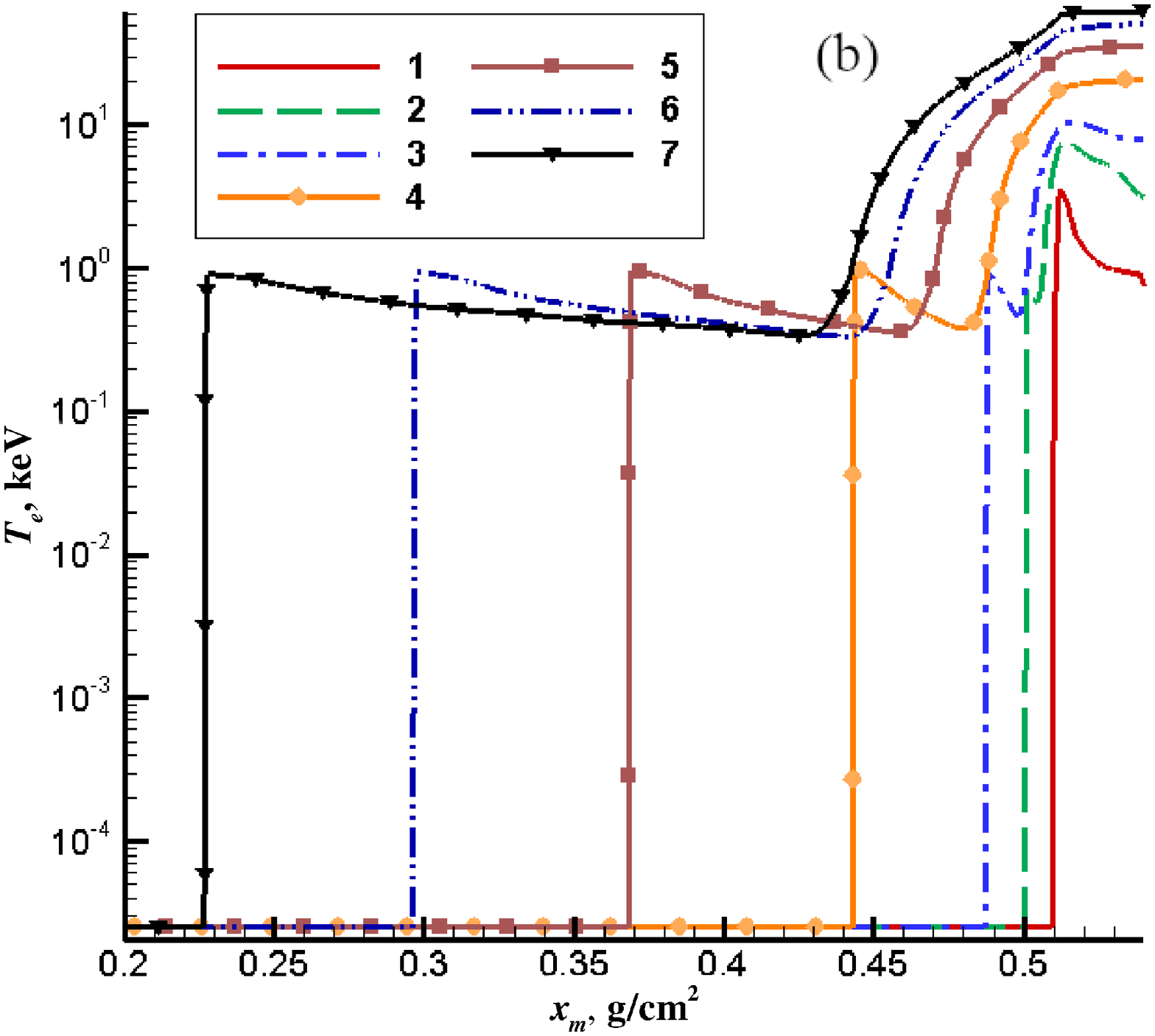}
    \end{minipage}
    \begin{minipage}[b]{0.45\textwidth}
        \includegraphics[width=\textwidth]{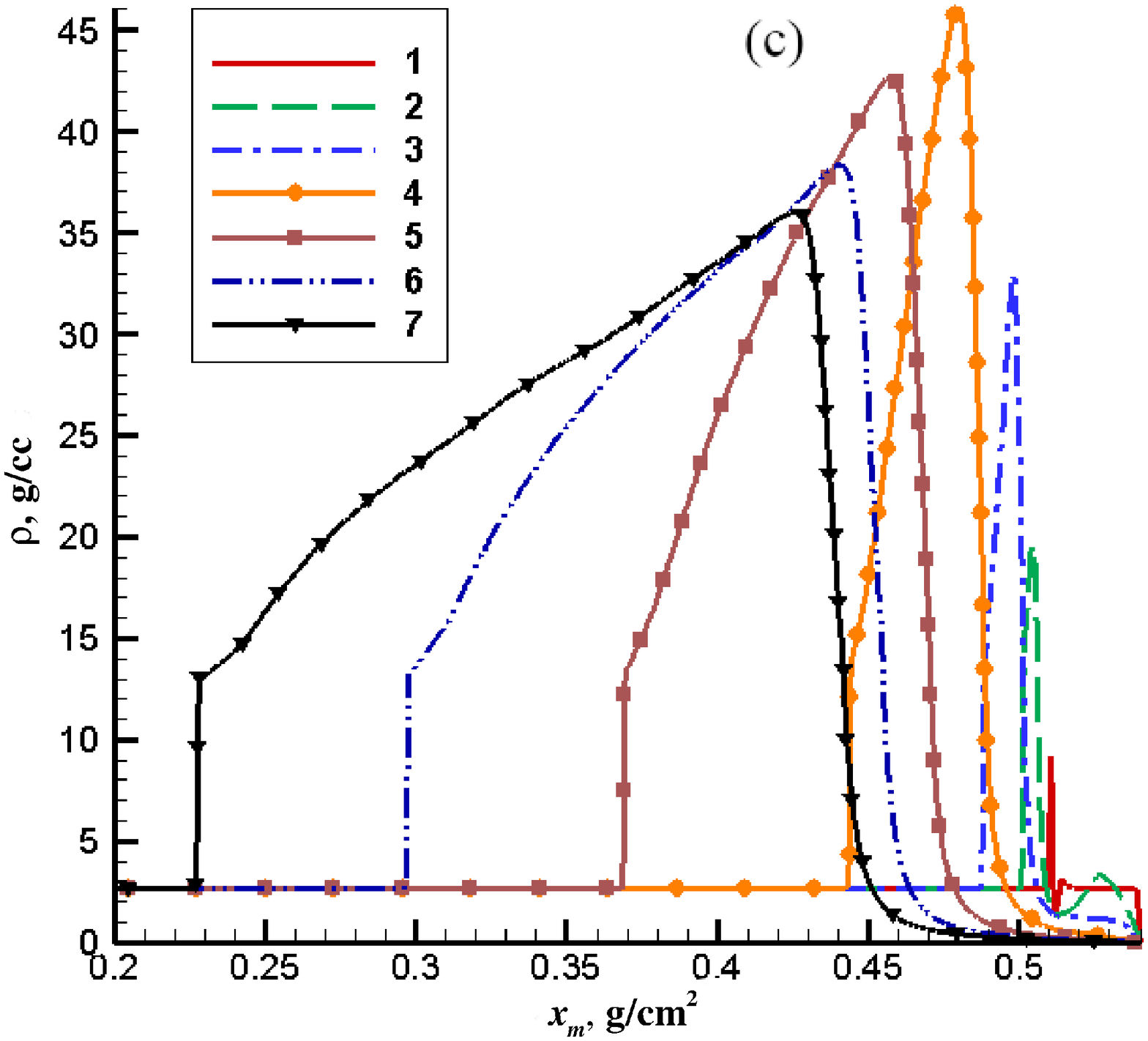}
    \end{minipage}
    \caption{Profiles of pressure (a), electron temperature (b) and density (c) over mass coordinate at various time moments: 20~ps (curves 1), 100~ps (2), 200~ps (3), 500~ps (4), 1~ns (5), 1.5~ns (6), 2~ns (7), obtained in the calculation under the impact of fast electron energy flux density $I_h = 2 \cdot 10^{17}$~W/cm$^{2}$ with initial energy of fast electrons $\varepsilon_h = 200$~keV.}\label{fig:001}
\end{figure}

\begin{figure}
  \includegraphics[width=0.6\textwidth]{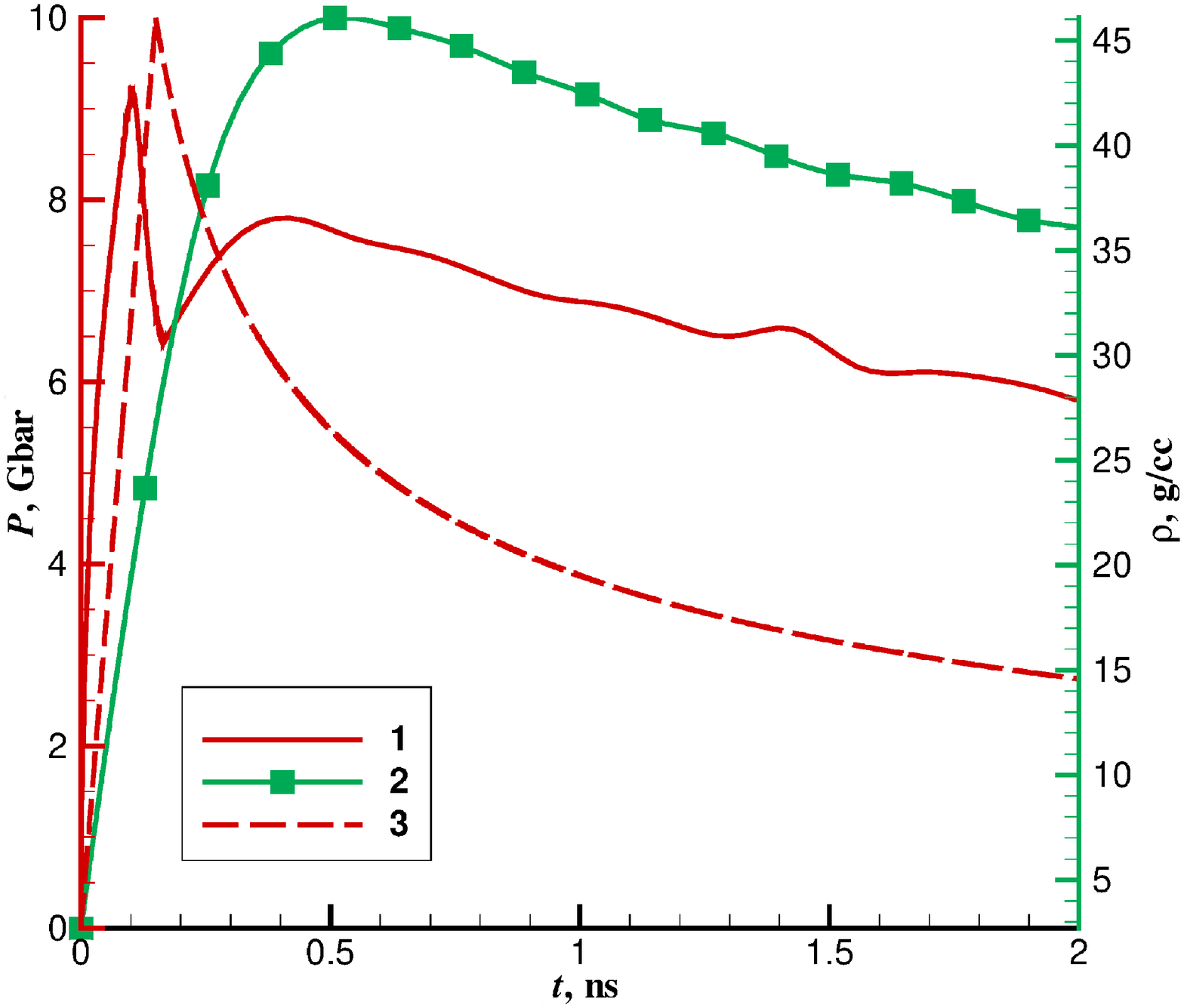}
  \caption{Time dependencies of the numerically calculated maximal values of pressure (curve 1) and density (2), and also a pressure in heated region according to analytical expressions~(\ref{eq:004})-(\ref{eq:007}) under the impact of fast electrons energy flux density $I_h = 2 \cdot 10^{17}$~W/cm$^{2}$ with initial energy of fast electrons $\varepsilon_h = 200$~keV.}\label{fig:002}
\end{figure}

\begin{figure}
  \includegraphics[width=0.6\textwidth]{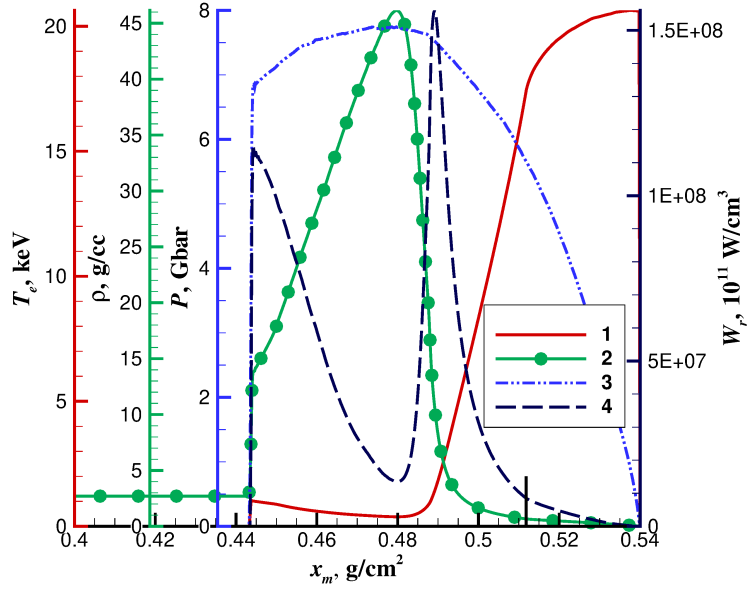}
  \caption{Profiles of electron temperature (1), density (2), pressure (3) and plasma emissivity (4) at time moment 500~ps, which corresponds to the achievement of maximal density in the peripheral region of shock wave. These data are obtained in calculation under the impact of fast electron energy flux density $I_h = 2 \cdot 10^{17}$~W/cm$^{2}$ with initial energy of fast electrons $\varepsilon_h = 200$~keV. Short vertical line indicates the inner boundary of heated region.\label{fig:003}}
\end{figure}

The bulk of numerical calculations was performed using the one-dimensional two-temperature hydrodynamic code DIANA~\cite{Zmitrenko1983}, supplemented by a module of energy transfer by fast electrons~\cite{Guskov2019}. DIANA code takes into account all the main relaxation and transport processes in the plasma and the real equations of states. The energy transfer by intrinsic plasma radiation is considered in approximation of volumetric losses due to bremsstrahlung mechanism. The cooling function operates by using threshold frequency for photons -- all photons with frequency above threshold value leave the plasma and take away their energy, in opposite case the energy is locally deposited. This code was successfully used for investigation of the effect of energy transfer by fast electrons on the gain of direct-drive ICF targets~\cite{Guskov2019,Guskov2019a}. The energy transfer from fast electrons in Coulomb collisions is calculated using the model of the stopping power of the plasma electrons adjusted for scattering by the plasma ions~\cite{Ribeyre2013,Atzeni2008,Honrubia2006}. The controlling calculations were performed using RADIAN numerical code~\cite{Vergunova1999,Rozanov2020}, which solves one-dimensional equations of two-temperature gas dynamics together with the radiation transfer equation. When solving the radiative transfer equation, the multi-group approximation, the method of characteristics, quasi-diffusion method and averaging over photon energies are used to effectively reduce dimension of equation. These calculations showed that taking radiation transfer into account led to insignificant deviations of thermodynamic parameters at the level of 5-10\% from the DIANA code results. Calculations using the DIANA code were performed in the range of fast electron energy flux density $I_h = 2 \cdot 10^{16}$--$2 \cdot 10^{19}$~W/cm$^2$, which, with the supposed laser energy conversion into fast electron energy $\eta = 0.2$, corresponds to the range of variation in laser intensity $I_L = 10^{17}$--$10^{20}$~W/cm$^2$. The initial energy of fast electrons was chosen according to scaling~(\ref{eq:001}) for the 1$^{\mbox{st}}$ harmonic of the Nd-laser radiation, and varied from a non-relativistic energy $\varepsilon_h = 100$~keV to a relativistic energy $\varepsilon_h = 3.8$~MeV. Figures~\ref{fig:001}-\ref{fig:003} show the results of a numerical calculation of shock wave propagation in an aluminum under a fast electron energy flux density of $I_h = 2 \cdot 10^{17}$~W/cm$^2$ and an initial energy $\varepsilon_h = 200$~keV (corresponds to the intensity of the 1$^{\mbox{st}}$ harmonic of Nd-laser radiation $I_L = 10^{18}$~W/cm$^2$ at the conversion $\eta = 0.2$). The mass range of fast electrons with indicated above energy is about of $\mu \approx 0.028$~g/cm$^2$. The mass range varies slightly with increasing temperature and decreasing density in the heated region in accordance with the change in the Coulomb logarithm. Figures~\ref{fig:001} show the distributions by the mass coordinate, respectively, of pressure, electron temperature, and density from the boundary of the semi-space, on which the source of fast electrons was set, to the shock wave front at the different time moments. The mass coordinates of the semi-space edge and  the inner boundary of the heated region are, respectively, 0.54~g/cm$^2$ and about of 0.512~g/cm$^2$. The figures show the time evolution of thermodynamic parameter distributions in the region heated by fast electrons and shock wave region. The pressure has a fairly uniform distribution behind the front of shock wave. Moreover, the pressure varies non-monotonically with time. At the initial time, the pressure behind the shock wave front reaches a value of about 10~Gbar at a time of about 120~ps, after which it decreases to a minimum of about 6.5~Gbar at a time of about 200~ps, then increases again reaching a second maximum of about 7.5~Gbar at a time of about 500~ps, and further decreases monotonously. The temperature distribution has a minimum, and the density distribution has a maximum in the peripheral region of the shock wave, whose positions fall on the same mass coordinate value. The maximum density in the radiation-cooled region reaches the value of about 45~g/cm$^3$ at the time moment of about 500~ps.
Figure~\ref{fig:002} shows the time dependencies of the numerically calculated values of maximum pressure and maximum density. This figure also shows the time dependence of the pressure in the heated region, calculated on the basis of the analytical solution~(\ref{eq:004})-(\ref{eq:007}) with a fraction of radiation energy loss of 40\%.
In the initial period of time up to 120-150~ps, when the initial stage of shock wave formation takes place, the maximum pressure is determined by the pressure in the heated region. Solution~(\ref{eq:004}) at the stage of quasi-static heating is in good agreement with the numerically calculated dependence. The pressure increases with time according to a law close to the linear one and reaches the values of about 10~Gbar by time moments of about 130-150~ps. The ablation loading time $t_h$  is about 150~ps, the maximum of the numerically calculated pressure is reached at $t = 130$~ps. During the ablation loading period the temperature behind the shock wave is about 1~keV and length of shock wave propagation is about 0.002~cm. Corresponding averaged mean free path of thermal photons is 0.01~cm, that is 5 times larger than the length of shock wave propagation. All these values are in a good agreement with the estimations of Section~\ref{sec:002} made on the base of expressions~(\ref{eq:004})--(\ref{eq:010}). As a result the whole shock wave is transparent and the growth of the density begins from the wave front and reaches the maximum in the peripheral back region. At subsequent time moments, the shock wave propagates under the conditions of expansion of the heated region as a whole and ablation of matter at a depth exceeding the mass range of fast electrons. This minimizes the numerically calculated pressure dependence at $t = 180$~ps. The formation of the steady motion of shock wave ends at a time of about 400~ps, after which the pressure decreases monotonically with time. The pressure in the steady-state shock wave is 30-40\% higher than the pressure of the analytical solution~(\ref{eq:005}) in the heated region, which is a consequence of hydrodynamic energy transfer to a substance of higher density. To the end of the period of formation of the steady-state movement of shock wave, maximum density values is achieved in its peripheral part. Subsequently, the maximum density value decreases in accordance with a decrease in pressure. Figure~\ref{fig:003} shows the profiles of pressure, density, electron temperature, and plasma emissivity at $t = 500$~ps, corresponding to the achievement of maximum density values in the peripheral part of shock wave. An increase in density in the peripheral region of shock wave takes place near the maximum of plasma emissivity. The maximum compression to a density of 45~g/cm$^3$ occurs in the region that is cooled to the maximum extent due to radiation energy losses under the action of the maximum pressure developed in the shock wave.
The degree of compression due to radiation cooling increases with laser pulse intensity, both due to an increase in pressure as well as an increase in radiation energy loss. In the calculation with impact of radiation of 1$^{\mbox{st}}$ harmonic of Nd-laser of the intensity $I_L = 10^{19}$~W/cm$^2$ ($\varepsilon_h = 1.2$~MeV) and laser energy conversion to fast electrons energy $\eta = 0.2$ the maximal density reaches the value of 65~g/cm$^3$.

Numerical simulations conducted for Cu target show the increase of the maximal density in peripheral area of the shock wave compared with an aluminum target. In the case of impact of the laser beam of the intensity $10^{18}$~W/cm$^2$ ($\varepsilon_h = 200$~keV, $\eta = 0.2$) the maximal density reaches the value about 58~g/cm$^3$, for the laser intensity $10^{19}$~W/cm$^2$ and the particles energy of 1.2~MeV it achieves the value of 66~g/cm$^3$. The saturation of the increase in density in the peripheral region of the shock wave with an increase in the atomic number is associated with saturation of the increase in radiation energy losses -- the increase in emissivity is compensated by a decrease in the transparency of the radiation region. In particular,  in the numerical simulation of Cu target with the laser intensity $10^{18}$~ W/cm$^2$ the time of ablation loading $t_h$ is about 60~ps and the temperature $T_h$ is about of 0.5~keV. The both values are approximately 2 times less, than in the case of Al target calculations, in accordance with~(\ref{eq:005}) and~(\ref{eq:006}). Corresponding averaged mean free path of thermal photons $l_r$ in the Cu-plasma with the density of, approximately, 3 times larger compared with Al-plasma is about 0.00006~cm. During the period of ablation loading the length of shock wave propagation is about 0.0005~cm that is 8 times larger than $l_r$. So, in contrast to the case of Al target, where whole shock wave is transparent and the growth of the density begins from the wave front and reaches the maximum in the peripheral back region, the main part of shock wave in Cu target is opaque and growth of the density occurs only in the narrow peripheral region. It should be noted, that for achieving higher densities the intensity of fast electron flux should be greater for the same initial energies of the particles but it is not supported yet by modern experiments, which now have a possibility to use the laser acceleration as a most powerful source of fast electrons.

The given calculation results allow one to determine the experimental conditions for studying the equation of state of matter. For a laser pulse with an intensity of $10^{18}$~W/cm$^2$, the pulse duration corresponding to the attainment of the maximum density values is about 100~ps. To generate a near plane shock wave, the radius of the laser beam must be at least two lengths of thermal expansion of the heating region, i.e. about 100 microns. This means that the energy of the laser pulse in this case should be about 4-5~kJ.

A shock wave generated by heating the solid by fast electrons may be of interest for studying the phenomena of radiative laboratory shock physics. The generation of such a wave occurs due to the work of a dense, high-temperature piston. As a result, a stratified object of strongly collisional plasma is formed (the Coulomb logarithm is about 10) with inhomogeneous distributions of thermodynamic parameters. The region of strong radiation, located in the peripheral region of the shock wave at the interface with the piston, is a relatively narrow region, the mass of which does not exceed 10\% of the mass of the piston. Electron thermal conductivity has an negligible effect on the thermodynamic state of this region. The region of radiation, as a plasma object formed by heating, is largely transparent to radiation. The optical thickness of the region heated by fast electrons (high-temperature piston) is 0.0001. The optical thickness in the region of strong radiation is about 0.1 for mid-Z  materials such as Al and about 0.5 for high-Z materials such as Cu. However, the optical thickness in the entire region covered by the shock wave can vary from fractions of a unit for mid-Z materials up to several units for high-Z materials. In the considered conditions of the problem, the P\'{e}clet number with respect to the electron thermal conductivity (the ratio of the hydrodynamic velocity to the velocity of the wave front of the electron heat conduction) for the radiation region is quite large for both Al and Cu targets, about 200-300. P\'{e}clet number in relation to radiation is different for different regions of discussed stratified object. In the region of strong radiation its value is about $10^6$ for Al target and $10^5$ for Cu target. The cooling parameter (ratio of cooling time and hydro time) is about 0.01 for mid-Z materials and about 0.1 for high-Z ones. Plasma objects with this kind of properties, formed by heating a substance with a petawatt flux of laser-accelerated fast electrons, can be of interest to laboratory astrophysics in connection with the controlled (due to a change in the laser pulse intensity and the selection of the target material) effect of radiation cooling.

\section{Conclusion}\label{sec:004}

Heating a solid with a beam of laser-accelerated fast electrons can provide the generation of a shock wave, which has characteristics unique for a laboratory experiment. The pressure behind the front of such a wave can reach several hundred and even thousands of Mbar at a temperature of several keV. Such a high temperature causes a high rate of radiation energy loss in a dense plasma, which is partially transparent to intrinsic radiation. As a result, conditions arise of strong radiation cooling and, as a consequence, of increasing in plasma compression in the peripheral part of shock wave. The performed theoretical and computational studies show that under the impact of radiation of the first harmonic of Nd-laser with intensity $10^{17}$--$10^{19}$~W/cm$^2$, the density of aluminum plasma in the peripheral region of fast-electron-driven shock wave can reach 30-60~g/cm$^3$ red and 50-70~g/cm$^3$ for copper plasma. With an increase in the atomic number of a substance, saturation of the increase in density occurs in the peripheral region of the shock wave, that is associated with saturation of the growth of radiation energy losses, at which the increase in emissivity is compensated by a decrease in the transparency of the radiation region. Investigation of the state of a substance at a pressure of several Gbar, temperature of several keV, and density of several tens of g/cm$^3$ represents a new section of the EOS study in a laboratory experiment. Investigation of the effect of increasing the density in the peripheral region of the shock wave, which is determined by radiative energy losses, is of great interest for establishing the optical properties of materials with a high atomic number. In addition, the dependence of the effect on the degree of ionization makes it possible to study the kinetics of ionization of matter at ultrahigh pressures, which is one of the fundamental questions of modern physics of high energy densities. Such experiments can be performed using a sub-nanosecond laser pulse with energy of 1-10~kJ.

%\begin{acknowledgments}
%This research was financially supported by the Russian Science Foundation under the project No.~16-11-10174.
%\end{acknowledgments}

\bibliography{refs}

\begin{thebibliography}{25}%
\makeatletter
\providecommand \@ifxundefined [1]{%
 \@ifx{#1\undefined}
}%
\providecommand \@ifnum [1]{%
 \ifnum #1\expandafter \@firstoftwo
 \else \expandafter \@secondoftwo
 \fi
}%
\providecommand \@ifx [1]{%
 \ifx #1\expandafter \@firstoftwo
 \else \expandafter \@secondoftwo
 \fi
}%
\providecommand \natexlab [1]{#1}%
\providecommand \enquote  [1]{``#1''}%
\providecommand \bibnamefont  [1]{#1}%
\providecommand \bibfnamefont [1]{#1}%
\providecommand \citenamefont [1]{#1}%
\providecommand \href@noop [0]{\@secondoftwo}%
\providecommand \href [0]{\begingroup \@sanitize@url \@href}%
\providecommand \@href[1]{\@@startlink{#1}\@@href}%
\providecommand \@@href[1]{\endgroup#1\@@endlink}%
\providecommand \@sanitize@url [0]{\catcode `\\12\catcode `\$12\catcode
  `\&12\catcode `\#12\catcode `\^12\catcode `\_12\catcode `\%12\relax}%
\providecommand \@@startlink[1]{}%
\providecommand \@@endlink[0]{}%
\providecommand \url  [0]{\begingroup\@sanitize@url \@url }%
\providecommand \@url [1]{\endgroup\@href {#1}{\urlprefix }}%
\providecommand \urlprefix  [0]{URL }%
\providecommand \Eprint [0]{\href }%
\providecommand \doibase [0]{http://dx.doi.org/}%
\providecommand \selectlanguage [0]{\@gobble}%
\providecommand \bibinfo  [0]{\@secondoftwo}%
\providecommand \bibfield  [0]{\@secondoftwo}%
\providecommand \translation [1]{[#1]}%
\providecommand \BibitemOpen [0]{}%
\providecommand \bibitemStop [0]{}%
\providecommand \bibitemNoStop [0]{.\EOS\space}%
\providecommand \EOS [0]{\spacefactor3000\relax}%
\providecommand \BibitemShut  [1]{\csname bibitem#1\endcsname}%
\let\auto@bib@innerbib\@empty
%</preamble>
\bibitem [{\citenamefont {Gus'kov}\ \emph {et~al.}(2012)\citenamefont
  {Gus'kov}, \citenamefont {Ribeyre}, \citenamefont {Touati}, \citenamefont
  {Feugeas}, \citenamefont {Nicolaï},\ and\ \citenamefont
  {Tikhonchuk}}]{Guskov2012}%
  \BibitemOpen
  \bibfield  {author} {\bibinfo {author} {\bibfnamefont {S.}~\bibnamefont
  {Gus'kov}}, \bibinfo {author} {\bibfnamefont {X.}~\bibnamefont {Ribeyre}},
  \bibinfo {author} {\bibfnamefont {M.}~\bibnamefont {Touati}}, \bibinfo
  {author} {\bibfnamefont {J.-L.}\ \bibnamefont {Feugeas}}, \bibinfo {author}
  {\bibfnamefont {P.}~\bibnamefont {Nicolaï}}, \ and\ \bibinfo {author}
  {\bibfnamefont {V.}~\bibnamefont {Tikhonchuk}},\ }\bibfield  {title}
  {\enquote {\bibinfo {title} {Ablation {P}ressure {D}riven by an {E}nergetic
  {E}lectron {B}eam in a {D}ense {P}lasma},}\ }\href {\doibase
  10.1103/physrevlett.109.255004} {\bibfield  {journal} {\bibinfo  {journal}
  {Physical Review Letters}\ }\textbf {\bibinfo {volume} {109}} (\bibinfo
  {year} {2012}),\ 10.1103/physrevlett.109.255004}\BibitemShut {NoStop}%
\bibitem [{\citenamefont {Ribeyre}\ \emph {et~al.}(2013)\citenamefont
  {Ribeyre}, \citenamefont {kov}, \citenamefont {Feugeas}, \citenamefont
  {Nicolaï},\ and\ \citenamefont {Tikhonchuk}}]{Ribeyre2013}%
  \BibitemOpen
  \bibfield  {author} {\bibinfo {author} {\bibfnamefont {X.}~\bibnamefont
  {Ribeyre}}, \bibinfo {author} {\bibfnamefont {S.~G.}\ \bibnamefont {kov}},
  \bibinfo {author} {\bibfnamefont {J.-L.}\ \bibnamefont {Feugeas}}, \bibinfo
  {author} {\bibfnamefont {P.}~\bibnamefont {Nicolaï}}, \ and\ \bibinfo
  {author} {\bibfnamefont {V.~T.}\ \bibnamefont {Tikhonchuk}},\ }\bibfield
  {title} {\enquote {\bibinfo {title} {Dense plasma heating and {G}bar shock
  formation by a high intensity flux of energetic electrons},}\ }\href
  {\doibase 10.1063/1.4811473} {\bibfield  {journal} {\bibinfo  {journal}
  {Physics of Plasmas}\ }\textbf {\bibinfo {volume} {20}},\ \bibinfo {pages}
  {062705} (\bibinfo {year} {2013})}\BibitemShut {NoStop}%
\bibitem [{\citenamefont {Gus'kov}(2014)}]{Guskov2014}%
  \BibitemOpen
  \bibfield  {author} {\bibinfo {author} {\bibfnamefont {S.~Y.}\ \bibnamefont
  {Gus'kov}},\ }\bibfield  {title} {\enquote {\bibinfo {title} {On the
  possibility of laboratory shock wave studies of the equation of state of a
  material at gigabar pressures with beams of laser-accelerated particles},}\
  }\href {\doibase 10.1134/s0021364014140069} {\bibfield  {journal} {\bibinfo
  {journal} {{JETP} Letters}\ }\textbf {\bibinfo {volume} {100}},\ \bibinfo
  {pages} {71--74} (\bibinfo {year} {2014})}\BibitemShut {NoStop}%
\bibitem [{\citenamefont {Cauble}\ \emph {et~al.}(1993)\citenamefont {Cauble},
  \citenamefont {Phillion}, \citenamefont {Hoover}, \citenamefont {Holmes},
  \citenamefont {Kilkenny},\ and\ \citenamefont {Lee}}]{Cauble1993}%
  \BibitemOpen
  \bibfield  {author} {\bibinfo {author} {\bibfnamefont {R.}~\bibnamefont
  {Cauble}}, \bibinfo {author} {\bibfnamefont {D.~W.}\ \bibnamefont
  {Phillion}}, \bibinfo {author} {\bibfnamefont {T.~J.}\ \bibnamefont
  {Hoover}}, \bibinfo {author} {\bibfnamefont {N.~C.}\ \bibnamefont {Holmes}},
  \bibinfo {author} {\bibfnamefont {J.~D.}\ \bibnamefont {Kilkenny}}, \ and\
  \bibinfo {author} {\bibfnamefont {R.~W.}\ \bibnamefont {Lee}},\ }\bibfield
  {title} {\enquote {\bibinfo {title} {Demonstration of 0.75 {G}bar planar
  shocks in x-ray driven colliding foils},}\ }\href {\doibase
  10.1103/physrevlett.70.2102} {\bibfield  {journal} {\bibinfo  {journal}
  {Physical Review Letters}\ }\textbf {\bibinfo {volume} {70}},\ \bibinfo
  {pages} {2102--2105} (\bibinfo {year} {1993})}\BibitemShut {NoStop}%
\bibitem [{\citenamefont {Karasik}\ \emph {et~al.}(2010)\citenamefont
  {Karasik}, \citenamefont {Weaver}, \citenamefont {Aglitskiy}, \citenamefont
  {Watari}, \citenamefont {Arikawa}, \citenamefont {Sakaiya}, \citenamefont
  {Oh}, \citenamefont {Velikovich}, \citenamefont {Zalesak}, \citenamefont
  {Bates}, \citenamefont {Obenschain}, \citenamefont {Schmitt}, \citenamefont
  {Murakami},\ and\ \citenamefont {Azechi}}]{Karasik2010}%
  \BibitemOpen
  \bibfield  {author} {\bibinfo {author} {\bibfnamefont {M.}~\bibnamefont
  {Karasik}}, \bibinfo {author} {\bibfnamefont {J.~L.}\ \bibnamefont {Weaver}},
  \bibinfo {author} {\bibfnamefont {Y.}~\bibnamefont {Aglitskiy}}, \bibinfo
  {author} {\bibfnamefont {T.}~\bibnamefont {Watari}}, \bibinfo {author}
  {\bibfnamefont {Y.}~\bibnamefont {Arikawa}}, \bibinfo {author} {\bibfnamefont
  {T.}~\bibnamefont {Sakaiya}}, \bibinfo {author} {\bibfnamefont
  {J.}~\bibnamefont {Oh}}, \bibinfo {author} {\bibfnamefont {A.~L.}\
  \bibnamefont {Velikovich}}, \bibinfo {author} {\bibfnamefont {S.~T.}\
  \bibnamefont {Zalesak}}, \bibinfo {author} {\bibfnamefont {J.~W.}\
  \bibnamefont {Bates}}, \bibinfo {author} {\bibfnamefont {S.~P.}\ \bibnamefont
  {Obenschain}}, \bibinfo {author} {\bibfnamefont {A.~J.}\ \bibnamefont
  {Schmitt}}, \bibinfo {author} {\bibfnamefont {M.}~\bibnamefont {Murakami}}, \
  and\ \bibinfo {author} {\bibfnamefont {H.}~\bibnamefont {Azechi}},\
  }\bibfield  {title} {\enquote {\bibinfo {title} {Acceleration to high
  velocities and heating by impact using {N}ike {KrF} laser},}\ }\href
  {\doibase 10.1063/1.3399786} {\bibfield  {journal} {\bibinfo  {journal}
  {Physics of Plasmas}\ }\textbf {\bibinfo {volume} {17}},\ \bibinfo {pages}
  {056317} (\bibinfo {year} {2010})}\BibitemShut {NoStop}%
\bibitem [{\citenamefont {Gus'kov}, \citenamefont {Zaretskii},\ and\
  \citenamefont {Kuchugov}(2020)}]{Guskov2020}%
  \BibitemOpen
  \bibfield  {author} {\bibinfo {author} {\bibfnamefont {S.~Y.}\ \bibnamefont
  {Gus'kov}}, \bibinfo {author} {\bibfnamefont {N.~P.}\ \bibnamefont
  {Zaretskii}}, \ and\ \bibinfo {author} {\bibfnamefont {P.~A.}\ \bibnamefont
  {Kuchugov}},\ }\bibfield  {title} {\enquote {\bibinfo {title} {Features and
  limiting characteristics of the heating of a substance by a laser-accelerated
  fast electron beam},}\ }\href {\doibase 10.1134/s0021364020030078} {\bibfield
   {journal} {\bibinfo  {journal} {{JETP} Letters}\ }\textbf {\bibinfo {volume}
  {111}},\ \bibinfo {pages} {135--138} (\bibinfo {year} {2020})}\BibitemShut
  {NoStop}%
\bibitem [{\citenamefont {Pease}(1957)}]{Pease1957}%
  \BibitemOpen
  \bibfield  {author} {\bibinfo {author} {\bibfnamefont {R.~S.}\ \bibnamefont
  {Pease}},\ }\bibfield  {title} {\enquote {\bibinfo {title} {Equilibrium
  {C}haracteristics of a {P}inched {G}as {D}ischarge {C}ooled by
  {B}remsstrahlung {R}adiation},}\ }\href {\doibase 10.1088/0370-1301/70/1/304}
  {\bibfield  {journal} {\bibinfo  {journal} {Proceedings of the Physical
  Society. Section B}\ }\textbf {\bibinfo {volume} {70}},\ \bibinfo {pages}
  {11--23} (\bibinfo {year} {1957})}\BibitemShut {NoStop}%
\bibitem [{\citenamefont {Vikhrev}(1978)}]{Vikhrev1978}%
  \BibitemOpen
  \bibfield  {author} {\bibinfo {author} {\bibfnamefont {V.~V.}\ \bibnamefont
  {Vikhrev}},\ }\bibfield  {title} {\enquote {\bibinfo {title} {Contraction of
  {Z}-pinch as a result of losses to radiation},}\ }\href@noop {} {\bibfield
  {journal} {\bibinfo  {journal} {Pis'ma Zh. Eksp. Teor. Fiz}\ }\textbf
  {\bibinfo {volume} {27}},\ \bibinfo {pages} {104--107} (\bibinfo {year}
  {1978})}\BibitemShut {NoStop}%
\bibitem [{\citenamefont {Bernal}\ and\ \citenamefont
  {Bruzzone}(2002)}]{Bernal2002}%
  \BibitemOpen
  \bibfield  {author} {\bibinfo {author} {\bibfnamefont {L.}~\bibnamefont
  {Bernal}}\ and\ \bibinfo {author} {\bibfnamefont {H.}~\bibnamefont
  {Bruzzone}},\ }\bibfield  {title} {\enquote {\bibinfo {title} {Radiative
  collapses in z-pinches with axial mass losses},}\ }\href {\doibase
  10.1088/0741-3335/44/2/306} {\bibfield  {journal} {\bibinfo  {journal}
  {Plasma Physics and Controlled Fusion}\ }\textbf {\bibinfo {volume} {44}},\
  \bibinfo {pages} {223--231} (\bibinfo {year} {2002})}\BibitemShut {NoStop}%
\bibitem [{\citenamefont {Craxton}\ \emph {et~al.}(2015)\citenamefont
  {Craxton}, \citenamefont {Anderson}, \citenamefont {Boehly}, \citenamefont
  {Goncharov}, \citenamefont {Harding}, \citenamefont {Knauer}, \citenamefont
  {McCrory}, \citenamefont {McKenty}, \citenamefont {Meyerhofer}, \citenamefont
  {Myatt}, \citenamefont {Schmitt}, \citenamefont {Sethian}, \citenamefont
  {Short}, \citenamefont {Skupsky}, \citenamefont {Theobald}, \citenamefont
  {Kruer}, \citenamefont {Tanaka}, \citenamefont {Betti}, \citenamefont
  {Collins}, \citenamefont {Delettrez}, \citenamefont {Hu}, \citenamefont
  {Marozas}, \citenamefont {Maximov}, \citenamefont {Michel}, \citenamefont
  {Radha}, \citenamefont {Regan}, \citenamefont {Sangster}, \citenamefont
  {Seka}, \citenamefont {Solodov}, \citenamefont {Soures}, \citenamefont
  {Stoeckl},\ and\ \citenamefont {Zuegel}}]{Craxton2015}%
  \BibitemOpen
  \bibfield  {author} {\bibinfo {author} {\bibfnamefont {R.~S.}\ \bibnamefont
  {Craxton}}, \bibinfo {author} {\bibfnamefont {K.~S.}\ \bibnamefont
  {Anderson}}, \bibinfo {author} {\bibfnamefont {T.~R.}\ \bibnamefont
  {Boehly}}, \bibinfo {author} {\bibfnamefont {V.~N.}\ \bibnamefont
  {Goncharov}}, \bibinfo {author} {\bibfnamefont {D.~R.}\ \bibnamefont
  {Harding}}, \bibinfo {author} {\bibfnamefont {J.~P.}\ \bibnamefont {Knauer}},
  \bibinfo {author} {\bibfnamefont {R.~L.}\ \bibnamefont {McCrory}}, \bibinfo
  {author} {\bibfnamefont {P.~W.}\ \bibnamefont {McKenty}}, \bibinfo {author}
  {\bibfnamefont {D.~D.}\ \bibnamefont {Meyerhofer}}, \bibinfo {author}
  {\bibfnamefont {J.~F.}\ \bibnamefont {Myatt}}, \bibinfo {author}
  {\bibfnamefont {A.~J.}\ \bibnamefont {Schmitt}}, \bibinfo {author}
  {\bibfnamefont {J.~D.}\ \bibnamefont {Sethian}}, \bibinfo {author}
  {\bibfnamefont {R.~W.}\ \bibnamefont {Short}}, \bibinfo {author}
  {\bibfnamefont {S.}~\bibnamefont {Skupsky}}, \bibinfo {author} {\bibfnamefont
  {W.}~\bibnamefont {Theobald}}, \bibinfo {author} {\bibfnamefont {W.~L.}\
  \bibnamefont {Kruer}}, \bibinfo {author} {\bibfnamefont {K.}~\bibnamefont
  {Tanaka}}, \bibinfo {author} {\bibfnamefont {R.}~\bibnamefont {Betti}},
  \bibinfo {author} {\bibfnamefont {T.~J.~B.}\ \bibnamefont {Collins}},
  \bibinfo {author} {\bibfnamefont {J.~A.}\ \bibnamefont {Delettrez}}, \bibinfo
  {author} {\bibfnamefont {S.~X.}\ \bibnamefont {Hu}}, \bibinfo {author}
  {\bibfnamefont {J.~A.}\ \bibnamefont {Marozas}}, \bibinfo {author}
  {\bibfnamefont {A.~V.}\ \bibnamefont {Maximov}}, \bibinfo {author}
  {\bibfnamefont {D.~T.}\ \bibnamefont {Michel}}, \bibinfo {author}
  {\bibfnamefont {P.~B.}\ \bibnamefont {Radha}}, \bibinfo {author}
  {\bibfnamefont {S.~P.}\ \bibnamefont {Regan}}, \bibinfo {author}
  {\bibfnamefont {T.~C.}\ \bibnamefont {Sangster}}, \bibinfo {author}
  {\bibfnamefont {W.}~\bibnamefont {Seka}}, \bibinfo {author} {\bibfnamefont
  {A.~A.}\ \bibnamefont {Solodov}}, \bibinfo {author} {\bibfnamefont {J.~M.}\
  \bibnamefont {Soures}}, \bibinfo {author} {\bibfnamefont {C.}~\bibnamefont
  {Stoeckl}}, \ and\ \bibinfo {author} {\bibfnamefont {J.~D.}\ \bibnamefont
  {Zuegel}},\ }\bibfield  {title} {\enquote {\bibinfo {title} {Direct-drive
  inertial confinement fusion: A review},}\ }\href {\doibase 10.1063/1.4934714}
  {\bibfield  {journal} {\bibinfo  {journal} {Physics of Plasmas}\ }\textbf
  {\bibinfo {volume} {22}},\ \bibinfo {pages} {110501} (\bibinfo {year}
  {2015})}\BibitemShut {NoStop}%
\bibitem [{\citenamefont {Golovkin}\ \emph {et~al.}(2006)\citenamefont
  {Golovkin}, \citenamefont {MacFarlane}, \citenamefont {Woodruff},
  \citenamefont {Bailey}, \citenamefont {Rochau}, \citenamefont {Peterson},
  \citenamefont {Mehlhorn},\ and\ \citenamefont {Mancini}}]{Golovkin2006}%
  \BibitemOpen
  \bibfield  {author} {\bibinfo {author} {\bibfnamefont {I.~E.}\ \bibnamefont
  {Golovkin}}, \bibinfo {author} {\bibfnamefont {J.~J.}\ \bibnamefont
  {MacFarlane}}, \bibinfo {author} {\bibfnamefont {P.~R.}\ \bibnamefont
  {Woodruff}}, \bibinfo {author} {\bibfnamefont {J.~E.}\ \bibnamefont
  {Bailey}}, \bibinfo {author} {\bibfnamefont {G.}~\bibnamefont {Rochau}},
  \bibinfo {author} {\bibfnamefont {K.}~\bibnamefont {Peterson}}, \bibinfo
  {author} {\bibfnamefont {T.~A.}\ \bibnamefont {Mehlhorn}}, \ and\ \bibinfo
  {author} {\bibfnamefont {R.~C.}\ \bibnamefont {Mancini}},\ }\bibfield
  {title} {\enquote {\bibinfo {title} {Spectroscopic analysis and {NLTE}
  radiative cooling effects in {ICF} capsule implosions with mid- dopants},}\
  }\href {\doibase 10.1016/j.jqsrt.2005.05.015} {\bibfield  {journal} {\bibinfo
   {journal} {Journal of Quantitative Spectroscopy and Radiative Transfer}\
  }\textbf {\bibinfo {volume} {99}},\ \bibinfo {pages} {199--208} (\bibinfo
  {year} {2006})}\BibitemShut {NoStop}%
\bibitem [{\citenamefont {Blondin}\ and\ \citenamefont
  {Cioffi}(1989)}]{Blondin1989}%
  \BibitemOpen
  \bibfield  {author} {\bibinfo {author} {\bibfnamefont {J.~M.}\ \bibnamefont
  {Blondin}}\ and\ \bibinfo {author} {\bibfnamefont {D.~F.}\ \bibnamefont
  {Cioffi}},\ }\bibfield  {title} {\enquote {\bibinfo {title} {The growth of
  density perturbations in radiative shocks},}\ }\href {\doibase
  10.1086/167955} {\bibfield  {journal} {\bibinfo  {journal} {The Astrophysical
  Journal}\ }\textbf {\bibinfo {volume} {345}},\ \bibinfo {pages} {853}
  (\bibinfo {year} {1989})}\BibitemShut {NoStop}%
\bibitem [{\citenamefont {Laming}(2004)}]{Laming2004}%
  \BibitemOpen
  \bibfield  {author} {\bibinfo {author} {\bibfnamefont {J.}~\bibnamefont
  {Laming}},\ }\bibfield  {title} {\enquote {\bibinfo {title} {Relationship
  between oscillatory thermal instability and dynamical thin-shell
  overstability of radiative shocks},}\ }\href {\doibase
  10.1103/physreve.70.057402} {\bibfield  {journal} {\bibinfo  {journal}
  {Physical Review E}\ }\textbf {\bibinfo {volume} {70}} (\bibinfo {year}
  {2004}),\ 10.1103/physreve.70.057402}\BibitemShut {NoStop}%
\bibitem [{\citenamefont {Beg}\ \emph {et~al.}(1997)\citenamefont {Beg},
  \citenamefont {Bell}, \citenamefont {Dangor}, \citenamefont {Danson},
  \citenamefont {Fews}, \citenamefont {Glinsky}, \citenamefont {Hammel},
  \citenamefont {Lee}, \citenamefont {Norreys},\ and\ \citenamefont
  {Tatarakis}}]{Beg1997}%
  \BibitemOpen
  \bibfield  {author} {\bibinfo {author} {\bibfnamefont {F.~N.}\ \bibnamefont
  {Beg}}, \bibinfo {author} {\bibfnamefont {A.~R.}\ \bibnamefont {Bell}},
  \bibinfo {author} {\bibfnamefont {A.~E.}\ \bibnamefont {Dangor}}, \bibinfo
  {author} {\bibfnamefont {C.~N.}\ \bibnamefont {Danson}}, \bibinfo {author}
  {\bibfnamefont {A.~P.}\ \bibnamefont {Fews}}, \bibinfo {author}
  {\bibfnamefont {M.~E.}\ \bibnamefont {Glinsky}}, \bibinfo {author}
  {\bibfnamefont {B.~A.}\ \bibnamefont {Hammel}}, \bibinfo {author}
  {\bibfnamefont {P.}~\bibnamefont {Lee}}, \bibinfo {author} {\bibfnamefont
  {P.~A.}\ \bibnamefont {Norreys}}, \ and\ \bibinfo {author} {\bibfnamefont
  {M.}~\bibnamefont {Tatarakis}},\ }\bibfield  {title} {\enquote {\bibinfo
  {title} {A study of picosecond laser{\textendash}solid interactions up to
  1019 w{\hspace{0.167em}}cm-2},}\ }\href {\doibase 10.1063/1.872103}
  {\bibfield  {journal} {\bibinfo  {journal} {Physics of Plasmas}\ }\textbf
  {\bibinfo {volume} {4}},\ \bibinfo {pages} {447--457} (\bibinfo {year}
  {1997})}\BibitemShut {NoStop}%
\bibitem [{\citenamefont {Haines}\ \emph {et~al.}(2009)\citenamefont {Haines},
  \citenamefont {Wei}, \citenamefont {Beg},\ and\ \citenamefont
  {Stephens}}]{Haines2009}%
  \BibitemOpen
  \bibfield  {author} {\bibinfo {author} {\bibfnamefont {M.~G.}\ \bibnamefont
  {Haines}}, \bibinfo {author} {\bibfnamefont {M.~S.}\ \bibnamefont {Wei}},
  \bibinfo {author} {\bibfnamefont {F.~N.}\ \bibnamefont {Beg}}, \ and\
  \bibinfo {author} {\bibfnamefont {R.~B.}\ \bibnamefont {Stephens}},\
  }\bibfield  {title} {\enquote {\bibinfo {title} {Hot-electron temperature and
  laser-light absorption in fast ignition},}\ }\href {\doibase
  10.1103/physrevlett.102.045008} {\bibfield  {journal} {\bibinfo  {journal}
  {Physical Review Letters}\ }\textbf {\bibinfo {volume} {102}} (\bibinfo
  {year} {2009}),\ 10.1103/physrevlett.102.045008}\BibitemShut {NoStop}%
\bibitem [{\citenamefont {Atzeni}, \citenamefont {Schiavi},\ and\ \citenamefont
  {Davies}(2008)}]{Atzeni2008}%
  \BibitemOpen
  \bibfield  {author} {\bibinfo {author} {\bibfnamefont {S.}~\bibnamefont
  {Atzeni}}, \bibinfo {author} {\bibfnamefont {A.}~\bibnamefont {Schiavi}}, \
  and\ \bibinfo {author} {\bibfnamefont {J.~R.}\ \bibnamefont {Davies}},\
  }\bibfield  {title} {\enquote {\bibinfo {title} {Stopping and scattering of
  relativistic electron beams in dense plasmas and requirements for fast
  ignition},}\ }\href {\doibase 10.1088/0741-3335/51/1/015016} {\bibfield
  {journal} {\bibinfo  {journal} {Plasma Physics and Controlled Fusion}\
  }\textbf {\bibinfo {volume} {51}},\ \bibinfo {pages} {015016} (\bibinfo
  {year} {2008})}\BibitemShut {NoStop}%
\bibitem [{\citenamefont {Honrubia}\ and\ \citenamefont {ter
  Vehn}(2006)}]{Honrubia2006}%
  \BibitemOpen
  \bibfield  {author} {\bibinfo {author} {\bibfnamefont {J.}~\bibnamefont
  {Honrubia}}\ and\ \bibinfo {author} {\bibfnamefont {J.~M.}\ \bibnamefont {ter
  Vehn}},\ }\bibfield  {title} {\enquote {\bibinfo {title} {Three-dimensional
  fast electron transport for ignition-scale inertial fusion capsules},}\
  }\href {\doibase 10.1088/0029-5515/46/11/l02} {\bibfield  {journal} {\bibinfo
   {journal} {Nuclear Fusion}\ }\textbf {\bibinfo {volume} {46}},\ \bibinfo
  {pages} {L25--L28} (\bibinfo {year} {2006})}\BibitemShut {NoStop}%
\bibitem [{\citenamefont {Afanasiev}\ and\ \citenamefont
  {Gus'kov}(1993)}]{Afanasiev1993}%
  \BibitemOpen
  \bibfield  {author} {\bibinfo {author} {\bibfnamefont {Y.~V.}\ \bibnamefont
  {Afanasiev}}\ and\ \bibinfo {author} {\bibfnamefont {S.~Y.}\ \bibnamefont
  {Gus'kov}},\ }\enquote {\bibinfo {title} {Nuclear {F}usion by {I}nertial
  {C}onfinement. {A} {C}omprehensive {T}reatise},}\ \ (\bibinfo  {publisher}
  {CRC Press},\ \bibinfo {year} {1993})\ Chap.\ \bibinfo {chapter} {Energy
  {T}ransfer to the {P}lasma in {L}aser {T}argets}, pp.\ \bibinfo {pages}
  {99--119}\BibitemShut {NoStop}%
\bibitem [{\citenamefont {Lindl}(1995)}]{Lindl1995}%
  \BibitemOpen
  \bibfield  {author} {\bibinfo {author} {\bibfnamefont {J.}~\bibnamefont
  {Lindl}},\ }\bibfield  {title} {\enquote {\bibinfo {title} {Development of
  the indirect-drive approach to inertial confinement fusion and the target
  physics basis for ignition and gain},}\ }\href {\doibase 10.1063/1.871025}
  {\bibfield  {journal} {\bibinfo  {journal} {Physics of Plasmas}\ }\textbf
  {\bibinfo {volume} {2}},\ \bibinfo {pages} {3933--4024} (\bibinfo {year}
  {1995})}\BibitemShut {NoStop}%
\bibitem [{\citenamefont {Zel'dovich}\ and\ \citenamefont
  {Raizer}(1967)}]{Zeldovich1967}%
  \BibitemOpen
  \bibfield  {author} {\bibinfo {author} {\bibfnamefont {Y.~B.}\ \bibnamefont
  {Zel'dovich}}\ and\ \bibinfo {author} {\bibfnamefont {Y.~P.}\ \bibnamefont
  {Raizer}},\ }\href@noop {} {\emph {\bibinfo {title} {Physics of shock waves
  and high-temperature hydrodynamic phenomena}}},\ edited by\ \bibinfo {editor}
  {\bibfnamefont {W.~D.}\ \bibnamefont {Hayes}}\ and\ \bibinfo {editor}
  {\bibfnamefont {R.~F.}\ \bibnamefont {Probstein}}\ (\bibinfo  {publisher}
  {Academic Press},\ \bibinfo {year} {1967})\BibitemShut {NoStop}%
\bibitem [{\citenamefont {Zmitrenko}\ \emph {et~al.}(1983)\citenamefont
  {Zmitrenko}, \citenamefont {Karpov}, \citenamefont {Fadeev}, \citenamefont
  {Shelaputin},\ and\ \citenamefont {Shpatakovskaya}}]{Zmitrenko1983}%
  \BibitemOpen
  \bibfield  {author} {\bibinfo {author} {\bibfnamefont {N.~V.}\ \bibnamefont
  {Zmitrenko}}, \bibinfo {author} {\bibfnamefont {V.~Y.}\ \bibnamefont
  {Karpov}}, \bibinfo {author} {\bibfnamefont {A.~P.}\ \bibnamefont {Fadeev}},
  \bibinfo {author} {\bibfnamefont {I.~I.}\ \bibnamefont {Shelaputin}}, \ and\
  \bibinfo {author} {\bibfnamefont {G.~V.}\ \bibnamefont {Shpatakovskaya}},\
  }\bibfield  {title} {\enquote {\bibinfo {title} {Description of the physical
  processes in the {DIANA} program for calculations of problems of laser
  fusion},}\ }\href@noop {} {\bibfield  {journal} {\bibinfo  {journal} {Voprosy
  Atomnoy Nauki i Tekhniki (VANT). Series Methods and Software for Numerical
  Solution of Problems of Mathematical Physics}\ ,\ \bibinfo {pages} {34--37}}
  (\bibinfo {year} {1983})}\BibitemShut {NoStop}%
\bibitem [{\citenamefont {Gus'kov}\ \emph
  {et~al.}(2019{\natexlab{a}})\citenamefont {Gus'kov}, \citenamefont
  {Kuchugov}, \citenamefont {Yakhin},\ and\ \citenamefont
  {Zmitrenko}}]{Guskov2019}%
  \BibitemOpen
  \bibfield  {author} {\bibinfo {author} {\bibfnamefont {S.~Y.}\ \bibnamefont
  {Gus'kov}}, \bibinfo {author} {\bibfnamefont {P.~A.}\ \bibnamefont
  {Kuchugov}}, \bibinfo {author} {\bibfnamefont {R.~A.}\ \bibnamefont
  {Yakhin}}, \ and\ \bibinfo {author} {\bibfnamefont {N.~V.}\ \bibnamefont
  {Zmitrenko}},\ }\bibfield  {title} {\enquote {\bibinfo {title} {Effect of
  `wandering' and other features of energy transfer by fast electrons in a
  direct-drive inertial confinement fusion target},}\ }\href {\doibase
  10.1088/1361-6587/ab0641} {\bibfield  {journal} {\bibinfo  {journal} {Plasma
  Physics and Controlled Fusion}\ }\textbf {\bibinfo {volume} {61}},\ \bibinfo
  {pages} {055003} (\bibinfo {year} {2019}{\natexlab{a}})}\BibitemShut
  {NoStop}%
\bibitem [{\citenamefont {Gus'kov}\ \emph
  {et~al.}(2019{\natexlab{b}})\citenamefont {Gus'kov}, \citenamefont
  {Kuchugov}, \citenamefont {Yakhin},\ and\ \citenamefont
  {Zmitrenko}}]{Guskov2019a}%
  \BibitemOpen
  \bibfield  {author} {\bibinfo {author} {\bibfnamefont {S.~Y.}\ \bibnamefont
  {Gus'kov}}, \bibinfo {author} {\bibfnamefont {P.~A.}\ \bibnamefont
  {Kuchugov}}, \bibinfo {author} {\bibfnamefont {R.~A.}\ \bibnamefont
  {Yakhin}}, \ and\ \bibinfo {author} {\bibfnamefont {N.~V.}\ \bibnamefont
  {Zmitrenko}},\ }\bibfield  {title} {\enquote {\bibinfo {title} {Effect of
  fast electrons on the gain of a direct-drive laser fusion target},}\ }\href
  {\doibase 10.1088/1361-6587/ab400e} {\bibfield  {journal} {\bibinfo
  {journal} {Plasma Physics and Controlled Fusion}\ }\textbf {\bibinfo {volume}
  {61}},\ \bibinfo {pages} {105014} (\bibinfo {year}
  {2019}{\natexlab{b}})}\BibitemShut {NoStop}%
\bibitem [{\citenamefont {Vergunova}\ and\ \citenamefont
  {Rozanov}(1999)}]{Vergunova1999}%
  \BibitemOpen
  \bibfield  {author} {\bibinfo {author} {\bibfnamefont {G.~A.}\ \bibnamefont
  {Vergunova}}\ and\ \bibinfo {author} {\bibfnamefont {V.~B.}\ \bibnamefont
  {Rozanov}},\ }\bibfield  {title} {\enquote {\bibinfo {title} {Influence of
  intrinsic {X}-ray emission on the processes in low-density laser targets},}\
  }\href {\doibase 10.1017/s0263034699173270} {\bibfield  {journal} {\bibinfo
  {journal} {Laser and Particle Beams}\ }\textbf {\bibinfo {volume} {17}},\
  \bibinfo {pages} {579--583} (\bibinfo {year} {1999})}\BibitemShut {NoStop}%
\bibitem [{\citenamefont {Rozanov}\ and\ \citenamefont
  {Vergunova}(2020)}]{Rozanov2020}%
  \BibitemOpen
  \bibfield  {author} {\bibinfo {author} {\bibfnamefont {V.~B.}\ \bibnamefont
  {Rozanov}}\ and\ \bibinfo {author} {\bibfnamefont {G.~A.}\ \bibnamefont
  {Vergunova}},\ }\bibfield  {title} {\enquote {\bibinfo {title} {Investigation
  of compression of indirect-drive targets under conditions of the {NIF}
  facility using one-dimensional modelling},}\ }\href {\doibase
  10.1070/qel17202} {\bibfield  {journal} {\bibinfo  {journal} {Quantum
  Electronics}\ }\textbf {\bibinfo {volume} {50}},\ \bibinfo {pages} {162--168}
  (\bibinfo {year} {2020})}\BibitemShut {NoStop}%
\end{thebibliography}%

\end{document}